\documentclass[jgrga]{agu2001}

\usepackage{amsmath}
\usepackage{amssymb}
\ifx\pdftexversion\undefined
\usepackage[dvips]{graphicx}
\else
\usepackage[pdftex]{graphicx}
\fi
\usepackage{psfrag}
\newcommand{\fig}{Fig.\hspace{0.05in}}

\authorrunninghead{LOBKOVSKY ET AL.}
\titlerunninghead{THRESHOLD PHENOMENA}

\authoraddr{Alexander E.~Lobkovsky, Department of Earth, Atmospheric
  and Planetary Sciences, \\ MIT, Cambridge, MA 02139, USA,
  \texttt{<leapfrog@mit.edu>}.}

\begin{document}

\articleid{1}{10}
\paperid{2004JF000172}
\journalid{109}{2004/12/24}

\author{Alexander E.~Lobkovsky}
\affil{Department of Earth, Atmospheric and Planetary Sciences, \\
  Massachusetts Institute of Technology, Cambridge, MA 02139, USA}

\author{Bill Jensen and Arshad Kudrolli} 
\affil{Department of Physics, Clark University, Worcester, MA 01610,
  USA}

\author{Daniel H.~Rothman}
\affil{Department of Earth, Atmospheric and Planetary Sciences, \\
  Massachusetts Institute of Technology, Cambridge, MA 02139, USA}

\vspace{0.2in}

\title{Threshold phenomena in erosion driven by subsurface flow}

\begin{abstract}
  We study channelization and slope destabilization driven by
  subsurface (groundwater) flow in a laboratory experiment.  The
  pressure of the water entering the sandpile from below as well as
  the slope of the sandpile are varied.  We present quantitative
  understanding of the three modes of sediment mobilization in this
  experiment: surface erosion, fluidization, and slumping.  The onset
  of erosion is controlled not only by shear stresses caused by
  surfical flows, but also hydrodynamic stresses deriving from
  subsurface flows.  These additional forces require modification of
  the critical Shields criterion.  Whereas surface flows alone can
  mobilize surface grains only when the water flux exceeds a
  threshold, subsurface flows cause this threshold to vanish at slopes
  steeper than a critical angle substantially smaller than the maximum
  angle of stability.  Slopes above this critical angle are unstable
  to channelization by any amount of fluid reaching the surface.
\end{abstract}

\begin{article}

\section{Introduction}
\label{sec:intro}

Unlike water, a layer of sand will not flow unless its surface is
inclined beyond a characteristic angle, known as the maximum angle of
stability \citep{duran99:_sands}.  This simple fact translates into a
host of threshold phenomena wherever granular material is found.  Many
such phenomena play a crucial role in the erosion of Earth's surface,
and very likely manifest themselves in the richness of the patterns
exhibited by drainage networks.
 
Depending on geological, hydrological, and climatological properties,
erosion by water is mainly driven either by overland flow or
subsurface flow.  The former case occurs when the shear stress imposed
by a sheet flow exceeds a threshold
\citep{horton45,loewenherz91,dietrich92,montgomery92,howard94}.
Erosion in the latter case---known as seepage erosion, or
sapping---occurs when a subsurface flow emerges on the surface.  Here
the eroding stresses derive not only from the resulting sheet flow but
also the process of seepage itself \citep{iverson86:_ground}.  The
onset of erosion for both overland flow and seepage is
threshold-dependent, but the additional source of stress in the case
of seepage has the potential to create significantly different erosive
dynamics.
 
Here we study the seepage case.  Whereas the case of Horton overland
flow has been extensively studied
\citep{smith72,dietrich92,montgomery94,dunne95,izumi95,izumi00:_linear},
seepage erosion has received less attention.  \cite{dunne80,dunne90}
suggests that erosive stresses due to seepage are more widespread in
typical environments than commonly assumed.  He also provides a
detailed description of seepage erosion in the field, together with a
discussion of the various factors that influence its occurrence.
Another focus of attention has been the controversial possibility that
many erosive features on Mars appear to have resulted from subsurface
flows
\citep{baker82:_mars,higgins82,laity85:_sapping,baker90:_spring,aharonson02}.
Although the importance of seepage stresses in erosion have been
realized by \cite{howard88b} and \cite{howard88}, comprehensive
quantitative understanding is difficult to obtain.  The complexity
arises from the interdependent motion of the sediment and fluid---the
``two-phase phenomenon'' \citep{yalin77:_mechanics}--- which, of
course, is common to {\em all} problems of erosion.
 
To further understand seepage erosion, we proceed from experiments
\citep{schumm87}.  Questions concerning the origin of ancient Martian
channels have motivated considerable experimental work in the past
\citep{kochel85,kochel86:_hawaii,howard88,kochel88}.  The process of
seepage erosion has also been studied as an example of drainage
network development \citep{gomez92}.  Our experiments, following those
of \cite{howard88,owoputi01:_erodability} and others, are designed to
enable us to construct a predictive, quantitative theory.
Consequently, they stress simplicity and completeness of information.
Although our setup greatly simplifies much of Nature's complexity, we
expect that at least some of our conclusions will improve general
understanding, and therefore be relevant to real, field-scale
problems.
 
A previous paper by \cite{schorghofer04:_spont} provided a qualitative
overview of the phenomenology in our experiment.  It described the
main modes of sediment mobilization: channelization, slumping, and
fluidization.  Here we provide quantitative understanding of the onset
and transitions between these modes.  

Our emphasis is on the threshold phenomena associated with the onset
of erosion, which we will ultimately characterize in the same way that
others \citep{duran99:_sands} have characterized the onset of dry
granular flow beyond the maximum angle of stability.  This involves a
construction of a generalized Shields criterion \citep{howard88} valid
in the presence of seepage through an inclined surface.  A major
conclusion is that the onset of erosion driven by seepage is
significantly different from the onset of erosion driven by overland
flow.  We find that there is a critical slope $s_\mathrm{c}$,
significantly smaller than the maximum angle of stability, above which
the threshold disappears.  Therefore any slope greater than
$s_\mathrm{c}$ is unstable to erosion if there is seepage through it.
This result is similar to well-known conclusions for the stability to
frictional failure of slopes with uniform seepage
\citep{taylor65:_fundamentals,iverson86:_ground,howard88}.  An
important distinction in our work, however, concerns the mode of
sediment mobilization and its local nature.  The existence of the
critical slope for seepage erosion may provide a useful quantitative
complement to the qualitative distinctions between seepage and
overland flow that have already been identified
\citep{laity85:_sapping}.

The remaining modes of sediment mobilization, fluidization and
slumping, are modeled using well established ideas
\citep{iverson86:_ground}.  The result of applying these ideas
together with the generalized Shields criterion provides a theoretical
prediction of the outcomes of the experiment, i.e., a phase diagram.
Agreement between theory and experiment is qualitative rather than
quantitative.  We nevertheless believe that our theoretical approach
is fundamentally sound and that better agreement would follow from
improved experimental procedures.

\section{Experiment}
\label{sec:expt}

\begin{figure}[htbp]
  \centering
  \includegraphics[width=3.3in]{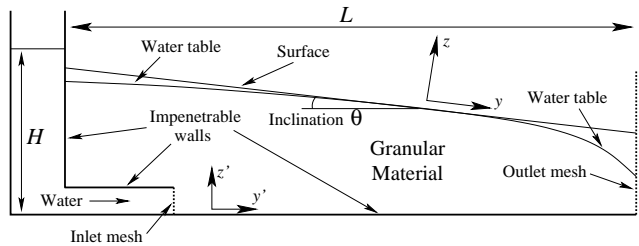}
  \caption{Schematic of the experiment as detailed in
    \cite{schorghofer04:_spont} as well as the setup for computing the
    bulk and surface water flow.}
  \label{fig:expt}
\end{figure}

In our experimental setup, first introduced by
\cite{schorghofer04:_spont}, a pile of identical cohesionless glass
beads $0.5$mm in diameter is saturated with water and compacted to
create the densest possible packing.  It is then shaped into a
trapezoidal wedge inclined at an angle $\theta$ with slope $s =
\tan\theta$ as shown in \fig\ref{fig:expt}.  The downslope length of
the wedge is $L = 90$ cm, its width across the slope is $119$ cm, and
its height in the middle is approximately $11$ cm.  Water enters the
sandpile underneath through a fine metal mesh and exits at the lower
end of the pile through the same kind of mesh.  A constant head at the
inlet is maintained by keeping a constant water level $H$ in the
reservoir behind the sandbox with the help of an outflow pipe.  The
slope $s$ of the pile and the water level $H$ are the control
parameters of the experiment.  The degree of packing of the granular
pile is the variable most difficult to control.

Our particular method of feeding water into the sandpile, similar to
that of \cite{owoputi01:_erodability}, can be motivated in three ways.
The most important justification is the fact that the amount of water
flowing on the surface can be finely controlled in our geometry.  This
feature is essential in probing the onset of erosion.  Second, our
setup allows us to access heads $H$ larger than the height of the
pile, which therefore allows us to explore dynamic regimes unavailable
if water enters the pile through a mesh in the back.  Third, a similar
seepage water flow geometry can exist in the field wherever water
travels beneath an impermeable layer that terminates.

\begin{figure}[htbp]
  \centering
  \includegraphics[width=3.3in]{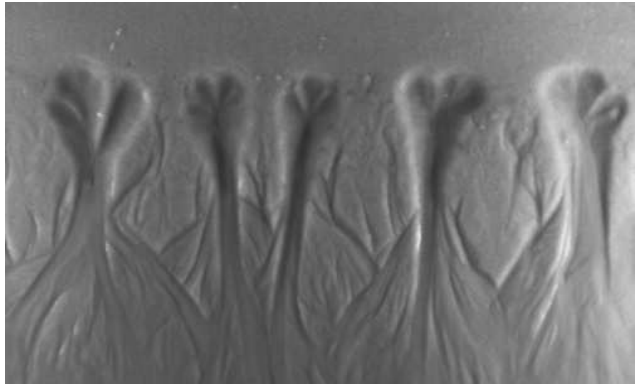}
  \caption{Examples of the channel network.  Top view of the slope of
    the eroding sandpile.  Branched channel heads migrate up the slope
    (upward) while the eroded sediment is deposited in the braided
    wash downslope.  The horizontal size of the image is approximately
    $1$m.  The slope of the pile is $s = 0.1$ and the water level $H =
    16.15$ cm.}
  \label{fig:channels}
\end{figure}

We have performed two types of experiments: steady and non-steady.
For a fixed water level and in absence of sediment motion, water flow
reaches steady state.  By monitoring the total water flux through the
system we estimate the time to reach steady state to be approximately
ten minutes.  To explore the onset of sediment motion, we raised the
water level $H$ in small increments, waiting each time for steady
state to be established.  Due to the particular shape of the bulk flow
in our experiment, surface flow exists over a finite region of the
surface.  The width of this seepage face and therefore the depth of
the surface flow can be tuned by changing $H$.  Because of the finite
extent of surface flow, its depth and therefore the viscous shear
stress reaches a maximum at a certain location.  Thus, by increasing
$H$ we can continuously tune the maximum shear stress experienced by
the surface grains.  The maximum shear stress reaches a critical value
for the onset of sediment motion in a certain location on the slope.
As we show below, we can compute where the maximum shear stress occurs
and thus can reliably detect the onset of sediment motion visually
because we focus our attention on this location.  Once sediment begins
to move, channels form almost immediately.  These channels grow in
length, width, and depth.  An example of the evolving channel network
is shown in \fig\ref{fig:channels}.  Depending on the slope, as the
channels deepen, the pile becomes unstable to fluidization or
slumping.  For slopes lower than approximately 0.05, the fluidization
threshold is reached before sediment is mobilized on the surface.

\begin{figure}[htbp]
  \centering
  \includegraphics[width=3.3in]{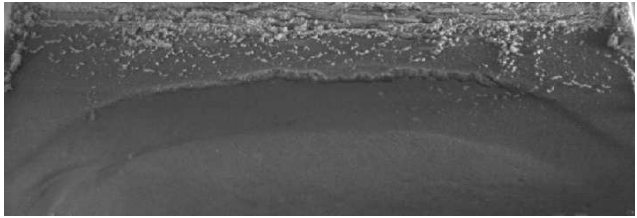}
  \caption{Example of a slumping sandpile viewed at an approximately
    45$^\circ$ angle to the slope after the water flow has been
    stopped.  The width of the imaged region is approximately $1$m.
    Slumping happens along a convex upward arc which looks darker
    because it is deeper and therefore wetter.}
  \label{fig:slump}
\end{figure}

We also explored the non-steady evolution of the bulk and surface
water flow and resulting sediment motion by raising the water level
$H$ to some higher value from zero.  In this case one of three things
can happen.  The pile can be fluidized within a few seconds or fail by
slumping as shown in \fig\ref{fig:slump}.  If this does not occur, the
water emerges on the surface just above the inlet.  A sheet of water
then washes down the slope of the pile.  During this initial wash,
sediment is mobilized and incipient channels form.  These channels
grow during subsequent relaxation of the bulk water flow towards
steady state.  Because of the initial wash's erosive power, channels
are able to form and grow for lower water pressures than in steady
experiments.

\begin{figure}[htbp]
  \centering
  \includegraphics[width=3.3in]{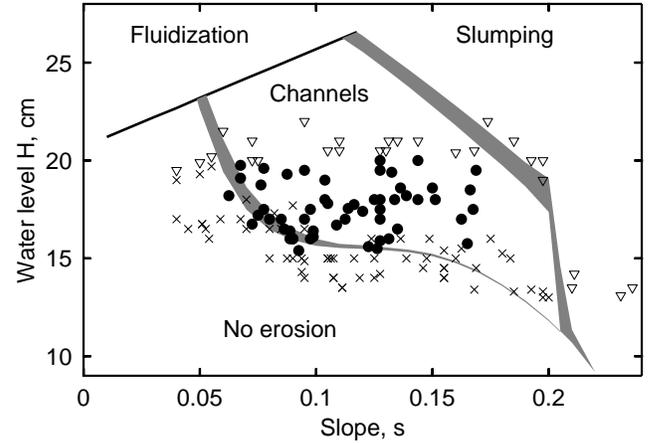}
  \caption{Experimental and theoretical phase diagram in the parameter
    space of slope and water level.  Experiments that yielded no
    erosion are denoted by $\times$; those that produced channels are
    indicated by $\bullet$; and those that produced fluidization
    and/or slumping within one hour of the beginning of the experiment
    are represented by $\triangledown$.  The straight line and
    gray-shaded curves are theoretical predictions for the boundaries
    separating the four regions indicated by their labels.  The
    thickness of the lines indicates uncertainty in the theory.  The
    boundary between the uneroded and channelized states is reasonably
    well approximated by our theory.  The theoretical boundaries for
    fluidization and slumping, however, appear to overestimate the
    critical water level, possibly as a result of inhomogeneities,
    dynamic changes in the sandpile's shape, or from the assumption of
    a steady state.}
  \label{fig:phase}
\end{figure}

Outcomes of a large number of non-steady experiments and several
steady experiments for varying slope $s$ and the water level $H$ are
summarized in the phase diagram in \fig\ref{fig:phase}.  Each symbol
in the plot represents one experiment.  The sediment is either
immobile (stable seepage), or it is mobilized on the surface where
channels form (channelization) or in the bulk (slumping or
fluidization).  In several experiments, slumping or fluidization
happened after channels formed and grew.  In the following sections we
describe the computations that allow us to construct the theoretical
boundaries between the three different modes of sediment mobilization
in our experiment.

\section{Calculation of the water flow}
\label{sec:flow}

Whereas steady-state flow can be readily characterized quantitatively,
non-steady flow characterization requires knowledge of the water-table
dynamics.  However, the theory of the water-table dynamics is less
well established than that of the flow through the bulk of a porous
medium.  Also, our steady-state experiments probe all aspects of
sediment dynamics.  We can therefore focus on the quantitative
characterization of the steady-state flow.

To study the onset of erosion quantitatively we need to be able to
establish a correspondence between the experimentally measurable
quantities such as the slope $s$, the water level $H$, the size of the
seepage face, and the water fluxes.  The seepage and surface fluxes
are the most difficult to measure.  In this section we set up their
computation.  The computation is designed to enable us to infer water
fluxes indirectly by measuring the size of the seepage face.  In the
following sections we will use this computation to quantify the onset
of erosion and to compute the slumping and liquefaction boundaries of
the channelization phase diagram shown in \fig\ref{fig:phase}.

\fig\ref{fig:expt} specifies the key quantities and coordinate systems
we use in computing the fluxes.  The flow profile is independent of
the $x$-coordinate across the slope of the sandpile except near the
side walls of the box.  We therefore treat the box as if it were
infinitely wide.  Flow is then two-dimensional and the specific
discharge vector $\mathbf{q}$ is in the $y$-$z$ plane.  We will use
two coordinate systems.  As shown in \fig\ref{fig:expt}, the $z'$
coordinate is measured vertically from the bottom of the box while the
$z$ coordinate is the normal distance away from the surface of the
pile.  The flow is governed by Darcy's law,
\begin{equation}
  \label{eq:darcy}
  \mathbf{q} = - K \nabla \psi,
\end{equation}
where $K$ is the scalar hydraulic conductivity, and 
\begin{equation}
  \label{eq:head}
  \psi = \pi + z'  
\end{equation}
is the total hydraulic head of the pore water.  Both $\mathbf{q}$ and
$K$ have units of velocity while the scaled pore pressure $\pi =
p/(\rho g)$ has units of length.  Here $\rho$ is the density of water
and $g$ is the magnitude of the acceleration of gravity.  We have
measured $K$ via a \textsf{U}-tube relaxation experiment.  To do so we
created a water level difference $\Delta h$ between the two arms of a
transparent \textsf{U}-shaped tube of width $\ell$ partially filled
with glass beads.  The rate of change of $\Delta h$ is given by
$K\Delta h/\ell$.  By measuring the rate of change of $\Delta h$ we
deduced the value of the hydraulic conductivity $K = 3.0 \pm 0.1$ mm/s
(\cite{schorghofer04:_spont}).  Hydraulic conductivity is sensitive to
the packing of the grains and is the variable most difficult to
control in our experiment.

Water incompressibility implies $\nabla \cdot \mathbf{q} = 0$,
therefore yielding Laplace's equation,
\begin{equation}
  \label{eq:laplace}
  \nabla^2 \pi = 0.
\end{equation}

To compute the pore pressure $\pi$, boundary conditions must be
specified.  The walls of the box are impenetrable.  Therefore the
discharge vector is parallel to the walls.  In other words, the flux
$q_\perp$ in the direction $\mathbf{\hat n}$ normal to the walls
vanishes.  Thus $q_\perp \equiv \mathbf{\hat n} \cdot \mathbf{q} = 0$.
Because the glass beads in our experiment are small, capillarity is
important.  When a tube filled with glass beads is lowered into a
reservoir of water, the porous bead-pack fully saturates in a region
that extends above the surface of the water by a capillary rise
$\pi_\mathrm{c}$.  We measured $\pi_\mathrm{c} = 25$ mm for our
material.  The capillary rise is a measure of the average radius of
the water menisci at the edge of the fully saturated zone.  The pore
pressure at the edge of the fully saturated zone is $-\pi_\mathrm{c}$
(without loss of generality we set the atmospheric pressure to zero).
Water can rise above the fully saturated zone through the smaller
pores and narrower throats.  Thus a partially saturated capillary
fringe exists above the fully saturated zone.  However, in this fringe
the water is effectively immobile since it is confined to the smaller
pores and narrower throats.

Since water flows only in the fully saturated zone, we define the
water table to be at its edge.  Thus, the pore pressure at the water
table is equal to the negative capillary rise $\pi = -\pi_\mathrm{c}$.
In steady state the discharge vector is parallel to the water table.
This extra condition allows us to determine the location of the water
table in steady state.

We neglect the pressure drop across the inlet mesh.  Therefore, the
pore pressure at the inlet mesh is $\pi = H - z'$.  The boundary
conditions at the surface of the sandpile and at the outlet mesh are
more subtle.  When no water seeps out, i.e., when the discharge vector
is parallel to the surface, the curvature of the water menisci between
grains can freely adjust so that the pressure $\pi$ can vary between
zero and $-\pi_\mathrm{c}$.  Therefore when $-\pi_\mathrm{c} < \pi <
0$, no seepage occurs.  Otherwise, the pore pressure equals the
atmospheric pressure $\pi = 0$ (we neglect the pressure exerted by the
thin layer of water on the surface), and the discharge vector has a
component normal to the surface, i.e., there is either exfiltration or
infiltration.

\begin{figure}[htbp]
  \centering
  \includegraphics[width=3.3in]{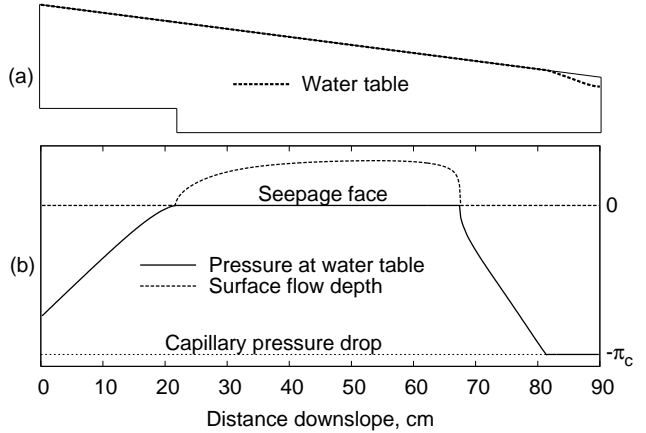}
  \caption{(a) Steady state location of the water table (dashed line)
    obtained by solving the Laplace's equation as detailed in the
    text.  Note that the water table is almost entirely at the surface
    of the sandpile except a for small region at the bottom of the
    sandpile.  (b) Pore pressure at the water table and surface flow.
    Vertical scale for the pressure at the water table is given by the
    negative capillary rise $-\pi_\mathrm{c}$, or the pressure at the
    water table when it is below the surface of the pile.  Note that
    seepage occurs only where the pore pressure reaches atmospheric
    pressure.  Slope is $s = 0.1$, water level $H = 15$ cm.}
  \label{fig:table_pressure_overland}
\end{figure}

To obtain the steady state location of the water table, we guess its
position and solve Laplace's equation~\eqref{eq:laplace} with the $\pi
= -\pi_\mathrm{c}$ boundary condition on the water table.  We then
move the water table in the direction of the local discharge vector by
an amount proportional to its length.  Iteration of this procedure
converges to the steady-state position of the water table.  An example
is shown in Figure \ref{fig:table_pressure_overland}.

Once the steady flow pattern is known, we can calculate the overland
water flux $Q(y)$ by integrating the one-dimensional continuity
condition which states that the downslope derivative of the overland
flux is equal to the seepage flux:
\begin{equation}
  \label{eq:continuity}
  \frac{dQ}{dy} = q_z \equiv \mathbf{\hat z} \cdot \mathbf{q} = -K
    \left.\frac{\partial \psi}{\partial z}\right|_{y=0}.
\end{equation}

\section{Onset of erosion and channel incision}
\label{sec:incision}

In this section we assume, based on direct observation, that the onset
of channel incision coincides with the onset of erosion (i.e., we
never observed a homogeneously eroding state).  In other words, when
the overland water flux becomes strong enough to carry grains, the
flow of sediment becomes immediately unstable to perturbations
transverse to the downslope direction and incipient channels form
\citep{smith72}.  Using this assumption and the calculation of the
overland water flux we can deduce the threshold condition for the
onset of erosion.

It is universally assumed after \cite{shields36} that the hydrodynamic
stresses exerted on the sandpile by the fluid flowing on its surface
determine whether cohesionless granular material is entrained.  In the
limit of laminar flow, the dominant hydrodynamic stress is the viscous
shear stress $\tau_\mathrm{b}$.  Appropriately scaled this shear
stress is termed the Shields number \citep{yalin77:_mechanics},
defined by
\begin{equation}
  \label{eq:shields_conv}
  \tau^* = \frac{\tau_\mathrm{b}}{(\rho_\mathrm{s} - \rho)gd},
\end{equation}
where $\rho_\mathrm{s}$ is the density of the granular material, $d$
is the grain diameter ($d = 0.5$ mm in our experiment), and the
surface is not inclined.

The conventional Shields number~\eqref{eq:shields_conv} is the ratio
between the horizontal force exerted by the flow and vertical force
due to grain's weight.  To generalize the notion of the Shields number
to the situation with seepage through an inclined surface, we make two
changes in Eq.~\eqref{eq:shields_conv}.  We first add the tangential
component of the seepage force density $\mathbf{f} = -\rho g \nabla
\psi$ acting over a length $d$ to the numerator
of~\eqref{eq:shields_conv}.  The numerator thus becomes
$\tau_\mathrm{b} + d (\mathbf{\hat y}\cdot \mathbf{f})$.  Note that we
did not include the tangential component of the grain's weight to the
numerator.  Defined in this way, the generalized Shields number
measures the effect of the fluid: both the bulk as well as the surface
flows.

\begin{figure}[htbp]
  \centering
  \includegraphics[width=2.3in]{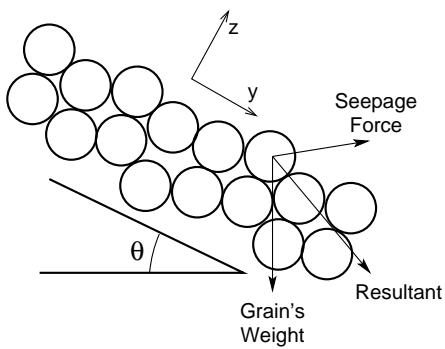}
  \caption{The denominator of the generalized Shields number is the
    projection onto the $z$-axis of the resultant of the grain's
    weight and the seepage force both scaled by $\pi d^2/6$.}
  \label{fig:mod_shields}
\end{figure}

Second, we replace the denominator of Eq.~\eqref{eq:shields_conv} by
the resultant (vectorial sum) of the seepage force on a grain and its
submerged weight, as shown in \fig\ref{fig:mod_shields}, both scaled
by $\pi d^2/6$ ($d^2$ to obtain stress as in the numerator and $\pi/6$
for agreement with the conventional Shields number), projected onto
the $z$-axis.  According to \cite{martin71:_seepage} the grains on the
surface of the bed experience a seepage force roughly half as large as
the grains several layers deep.  Consequently, we assume that the
seepage force is reduced by a factor of $a \approx 0.5$; therefore
\begin{equation}
  \label{eq:shields}
  \tau^* = \frac{\tau_\mathrm{b} - a\rho g d \, (\partial \psi/\partial
  y)}{(\rho_\mathrm{s} - \rho)gd\cos\theta + a\rho g d \,
    (\partial \psi/\partial z)},
\end{equation}
where $\theta$ is the inclination angle of the surface.  The
importance of the seepage stresses for the criterion for the onset of
erosion was previously realized by \cite{howard88}.  It can be shown
that their equation (10) expressing marginal stability of a surface
grain is equivalent to writing $\tau^* = \tan\phi$, the tangent of the
angle of internal friction.

The generalized Shields number Eq.~\eqref{eq:shields} is a measure of the
relative importance of the tangential and normal forces acting on a
grain at the surface of the sandpile.  Therefore, we expect $\tau^*$
to be a control parameter for erosion.  In other words, there exists a
critical Shields number $\tau^*_\mathrm{c}$, such that when $\tau^* <
\tau^*_\mathrm{c}$, surface grains are immobile, and when $\tau^* >
\tau^*_\mathrm{c}$, sediment is mobilized.  Note that
(\ref{eq:shields}) reduces to the classical definition of the Shields
number for a flat surface without seepage.  Also note that since we
did not include the tangential component of grain's weight, the
critical Shields number at the onset of sediment motion vanishes when
the inclination angle $\theta$ reaches the maximum angle of stability.

Although we obtain the seepage force density $\mathbf{f}$ as a result
of computing the pore-water pressure, to calculate the boundary shear
stress $\tau_\mathrm{b}$ we must estimate the thickness of the surface
water layer $h(y)$.  Since this thickness changes slowly in the
downslope direction, we can approximate the surface flow by the steady
flow of a uniform layer of viscous fluid.  Also, the surface water
flux is small enough for turbulence to be of no importance.  The
thickness $h$ of laminar surface flow for a given flux $Q$ is
\citep{landau87:_fluid}
\begin{equation}
  \label{eq:thickness}
    h = \left(
      \frac{3Q \, \eta}{\rho g \sin\theta}
    \right)^{1/3},
\end{equation}
where $\eta$ is the viscosity, while the viscous shear stress exerted
on the sandpile is
\begin{equation}
  \label{eq:tau}
  \tau_\mathrm{b} = \rho g h \sin\theta.
\end{equation}
The particle Reynolds number can then be calculated using the bottom
shear stress~\eqref{eq:tau} and shear velocity $v^* =
\sqrt{\tau_\mathrm{b}/\rho}$ as
\begin{equation}
  \label{eq:Re}
  \mathrm{Re} = \frac{v^* d}{\nu},
\end{equation}
where $\nu =\eta/\rho$ is the kinematic viscosity of water.  We
estimate that in our experiments, this particle Reynolds number varies
between 5 and 20 depending on the slope $s$ of the pile and the water
level $H$.  We verify this estimate of the Reynolds number by a direct
measurement of the thickness of the surface flow.  We find that this
thickness is several grain diameters.  This justifies the laminar flow
assumption used in obtaining Eq.~\eqref{eq:thickness}.

Using~\eqref{eq:tau}, the Shields number~\eqref{eq:shields} can now
be conveniently rewritten as
\begin{equation}
  \label{eq:shields-prefinal}
  \tau^* = \frac{(h/d)\sin\theta - a \, (\partial \psi/\partial
    y)}{(\rho_\mathrm{s}/\rho - 1) \cos\theta + a \, (\partial
    \psi/\partial z)}. 
\end{equation}
This expression can be further simplified by noting that $\pi = 0$
along the seepage face.  Therefore $\partial \psi/\partial y =
-\sin\theta$ at the surface wherever there is overland flow.  We
arrive at the final expression for the modified Shields number which
depends on the surface flow thickness $h$, the normal component
$\partial \psi/\partial z$ of the seepage force density at the
surface, and the seepage force reduction factor $a$
\begin{equation}
  \label{eq:shields-prefinal1}
  \tau^* = \frac{(h/d + a)\sin\theta}{(\rho_\mathrm{s}/\rho - 1)
    \cos\theta + a \, (\partial \psi/\partial z)}.
\end{equation}

In our geometry, both the surface and the seepage water fluxes reach a
maximum somewhere along the slope.  Therefore the Shields number has a
maximum value as well.  Below we calculate this maximum Shields number
in steady state for a given slope $s$ and water level $H$.

\section{Critical slope for the onset of seepage erosion}
\label{sec:crit-slope-seep}

\begin{figure}[htbp]
  \centering
  \includegraphics[width=3.3in]{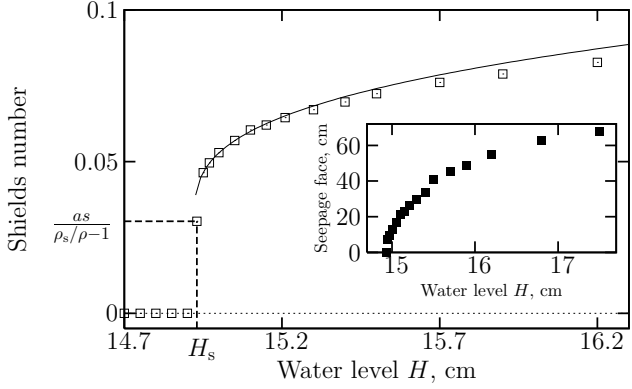}
  \caption{Computed maximum Shields number versus the water level $H$
    for a pile of slope $s = 0.1$.  At $H = H_\mathrm{s}$ water first
    seeps through the surface and the Shields number jumps to a
    nonzero value.  Afterwards it increases rapidly as $(H -
    H_\mathrm{s})^{1/3}$ (solid line).  Inset: corresponding size of
    the computed seepage face.}
  \label{fig:shields_H}
\end{figure}

We now explore the consequences of seepage for the phenomenology of
the onset of erosion.  Because of the additional force on the surface
grains, seepage flow is more erosive than overland flow.  This notion
is reflected quantitatively in the generalized Shields number.  Let us
examine how the maximum Shields number varies with the water level $H$
in our experiment.  A representative graph of the maximum Shields
number versus the water level is shown in \fig\ref{fig:shields_H}.
Below a water level $H_\mathrm{s}(s)$ that is a function of the slope,
no water seeps out to the surface of the pile.  Even though the water
table may be at the surface, the pressure at the water table is below
atmospheric pressure and capillarity prevents seepage.  When $H =
H_\mathrm{s}$, i.e., exactly at the onset of seepage, the pressure
reaches $\pi = 0$ at some point on the surface.  Since the seepage
flux is still zero, $\partial \psi/\partial z = 0$ along the wet part
of the surface.  Therefore, just above the seepage onset, when the
water layer thickness and the seepage flux are both infinitesimally
small, the maximum Shields number is
\begin{equation}
  \label{eq:shields_onset}
  \tau^*_\mathrm{s}(s) = \frac{as}{\rho_\mathrm{s}/\rho - 1}.
\end{equation}
In contrast to overland flow, the consequence of seepage is that as
soon as the water emerges on the surface, the maximum Shields number
is some non-zero value which depends on the slope.  This also implies
that there exists a critical slope $s_\mathrm{c}$ such that
$\tau_\mathrm{s}(s_\mathrm{c})$ is equal to the critical Shields
number $\tau^*_\mathrm{c}$, i.e.,
\begin{equation}
  \label{eq:crit_slope}
  s_\mathrm{c} = \tau^*_\mathrm{c}(\rho_\mathrm{s}/\rho - 1)/a.
\end{equation}
For slopes greater than $s_\mathrm{c}$ seepage is always erosive.
Note that for low-density particles this critical slope can be
arbitrarily small.  The expression for the critical slope for seepage
erosion in Eq.~\eqref{eq:crit_slope} is analogous to well-known
formulas for stability of slopes to Coulomb failure due to uniform
seepage \citep{iverson86:_ground}.  Our result applies locally to the
point where non-uniform seepage first emerges on the surface.  In this
situation, the pile is generally stable to Coulomb failure and the
sediment is eroded only locally on the surface.

\begin{figure}[htbp]
  \centering
  \includegraphics[width=2.3in]{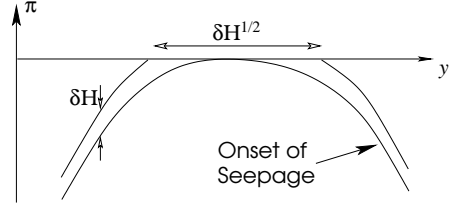}
  \caption{Schematic of the pore pressure $\pi$ as a function of the
    downslope coordinate $y$ at the onset of seepage $H =
    H_\mathrm{s}$ and just above.}
  \label{fig:increment}
\end{figure}

We now show that above $H_\mathrm{s}$, the maximum Shields number
increases rapidly as a $1/3$ power of the water level excess $(H -
H_\mathrm{s})^{1/3}$.  At the onset of seepage, i.e., when $H =
H_\mathrm{s}$, the pressure at the water table reaches atmospheric
pressure $\pi = 0$ at some point $P$ located at $z = 0$ and $y =
y_\mathrm{s}$ on the surface.  Even though the water table is at the
surface, there is no seepage anywhere, i.e., $\partial \psi/\partial z
= 0$.  Because the pressure is smooth, it can be approximated by a
quadratic function near this point so that $\pi \approx b z^2 - c(y -
y_\mathrm{s})^2$, where $b$ and $c$ are constants with appropriate
dimensions.  When the water level is raised by a small increment
$\delta H = H - H_\mathrm{s}$, the lowest order change in the head at
the water table is an increase of $\delta H$ with the exception of the
region where this increase would lead to a positive pressure.  As
illustrated in \fig\ref{fig:increment}, in this region the pore
pressure $\pi$ is set to $0$, and thus this region becomes the seepage
face.  The width of the seepage face $W$ scales like the square root
of $\delta H$, i.e., $W \sim \delta H^{1/2}$ as seen in the inset of
Fig.\hspace{0.05in}\ref{fig:shields_H}.  The seepage flux can be
estimated by noting that the hydraulic head is modified by an amount
$\delta H$ over a vertical region of order $W$.  Therefore we obtain
$\partial \psi/\partial z \sim \delta H /W \sim \delta H^{1/2}$.  The
total surface flux therefore scales like the product of the seepage
flux and the width $W$ of the seepage face, i.e., $Q \sim W \,\partial
\psi/\partial z \sim \delta H$.  The lowest order change in the
maximum Shields number is due to the change of the surface flow depth
$h \sim Q^{1/3} \sim \delta H^{1/3}$.  Thus as we claimed above, just
above the water level $H_\mathrm{s}$ for the onset of seepage,
\begin{equation}
  \label{eq:shields_seepage}
  \tau^* \approx \tau^*_\mathrm{s}(s) + \alpha(s) (H - H_\mathrm{s})^{1/3},
\end{equation}
where the constant $\alpha$ is a function of the slope.  Variation
with water level of the computed maximum Shields number shown in
\fig\ref{fig:shields_H} is consistent with
Eq.~\eqref{eq:shields_seepage}.

\section{Measurement of the critical Shields number}
\label{sec:meas-crit-shields}

In the previous sections we have detailed the way of calculating the
bulk and surface water fluxes in our experiment and the resulting
maximum generalized Shields number.  In this section we use this
calculation to examine the onset of the sediment flow and
channelization.  Our first goal is to measure the threshold or
critical Shields number required for the mobilization of sediment.  We
then use this measured value of the critical Shields number to predict
the outcome of steady-state experiments for various values of the
slope and the water level and thus compute the channelization boundary
in the phase diagram in \fig\ref{fig:phase}.

The actual maximum Shields number in the experiment differs from the
quantity calculated in Eq.~(\ref{eq:shields-prefinal1}).  In addition
to random errors in the measurements of the pile dimensions and water
level, there are several sources of systematic error.  For example,
the pressure drop across the inlet mesh results in a lower effective
hydraulic head.  Also, our measurement of the capillary rise
$\pi_\mathrm{c}$ is dependent on a visual estimate of the fully
saturated zone and thus can be a source of systematic error.  We
indeed find that the size of the seepage face calculated for a
particular water level $H$ is greater than measured in the experiment.
However, the size of the seepage face translates directly into the
surface water flux and therefore the maximum Shields number.

\begin{figure}[htbp]
  \centering
  \includegraphics[width=3.3in]{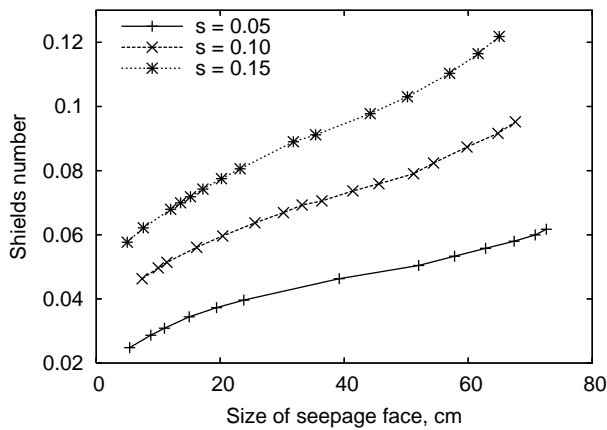}
  \caption{Computed maximum Shields number versus the seepage face
    size for three different slopes.  We use this calculation to infer
    the Shields number from the experimentally measured size of the
    seepage face.}
  \label{fig:shields_seep}
\end{figure}

The inset of Figure \ref{fig:shields_H} shows the typical dependence
of the size of the seepage face on the water level.  The variation of
the maximum Shields number with the size of the seepage face is shown
in \fig\ref{fig:shields_seep} for three different slopes.  We use
this computed correspondence between the size of the seepage face and
the maximum Shields number to infer the maximum Shields number in the
experiment by measuring the size of the seepage face.

To measure the critical Shields number we raise the water level $H$ in
increments of a few millimeters at a time.  Each time the water level
is increased, the seepage flow is allowed to reach a steady state.  In
each of these steady states we measure the seepage face size and infer
the corresponding maximum Shields number.  Eventually, sediment is
mobilized and we record the size of the seepage face and compute the
corresponding maximum Shields number.  This number is an upper bound
on the critical Shields number for our granular material at that
particular slope.  The lower bound on the critical Shields number is
obtained from the largest seepage face at which no sediment is moving
or sediment motion is only transient.  Averaging over several
experiments with slope $s = 0.1$ we estimate the critical Shields
number to be
\begin{equation}
  \label{eq:Shields_crit}
  \tau^*_\mathrm{c} = 0.085 \pm 0.005.
\end{equation}
It is not obvious that the generalized critical Shields number for the
onset of seepage driven erosion should coincide with the critical
Shields number for overland flow.  However our measured value of the
critical generalized Shields number is within the scatter of the
existing data for overland flow summarized in
\cite{buffington97:_systematic}.  Our measurement the critical
generalized Shields number is equivalent to measuring the angle of
internal friction due to the correspondence of our definition of
$\tau^*$ \eqref{eq:shields} and Howard and McLane's equation (10).

Deviations from flatness of the pile's surface result in the
fluctuations of the thickness of the surface water film.  As a result,
the maximum bottom shear stress in the experiment is systematically
greater than that calculated at a given size of the seepage face.
Thus the Shields number calculated for a particular size of the
seepage face is the lower bound on the actual Shields number in the
experiment.

In principle, the critical Shields number should vary with the slope
of the pile.  Evidence for this is the fact that at the maximum angle
of stability any additional forcing from the water flowing over the
bed mobilizes sediment.  Since it is reasonable to assume that the
critical Shields number is continuous and monotonic, we arrive at the
notion that it decreases monotonically with slope and vanishes at the
maximum angle of stability.  For small slopes the critical Shields
number is expected to decrease as the cosine of the inclination angle
since this is the lowest order change in the stabilizing effect of
gravity.  For most slopes in our experiments, $\cos\theta$ is within a
few percent of unity and thus we can ignore the variation of the
critical Shields number with slope.

This assumption allows us to predict the water level at which erosion
and therefore channelization should commence in our experiment.  In
Fig.\hspace{0.05in}\ref{fig:phase} a boundary is drawn between regions
where sediment is expected to mobilize and remain immobile.  To obtain
this line we computed for each slope the water level $H_\mathrm{e}(s)$
at which the Shields number is equal to the critical Shields number
\eqref{eq:Shields_crit}.  Below this water level, i.e., when $H <
H_\mathrm{e}(s)$, the maximum Shields number is below critical and
thus sediment is immobile.  Conversely, for $H > H_\mathrm{e}(s)$, the
maximum Shields number is above critical and thus sediment is
mobilized and channels form.  The channelization boundary is widened
because of the uncertainty in the critical Shields number.

Qualitative agreement of the channelization boundary with experiments
is perhaps due to the opposite action of two effects.  First,
channelization occurs for lower water levels in non-steady
experiments.  This happens because in non-steady experiments the
maximum Shields number overshoots its steady-state value.  The
overshoot is greatest for small slopes.  Second, a pressure drop
across the inlet mesh and the compacted region of sand close to it has
an opposite effect which increases the water level needed for
channelization.  These two effects, though small, could together
affect the accuracy of our predictions.  Since these effects act in
opposite ways, our the predictions of the calculated channelization
water level $H_\mathrm{e}(s)$ agree qualitatively with the
experiments.

\section{Fluidization and slumping}
\label{sec:fluid}

Having computed the channelization boundary in the phase diagram, we
now pursue a quantitative description of the other two modes of
sediment mobilization exhibited by our sandpile.  Higher water
pressures can cause the sandpile to fail in one of two ways.  First,
an upward seepage force can lift sand and result in a fluidization or
quicksand instability.  Second, the pile can become unstable to
slipping, slumping, or sliding.  Both failure mechanisms have been
discussed by a number of studies, e.g., those of
\cite{iverson86:_ground} or \cite{martin71:_seepage}.

\begin{figure}[htbp]
  \centering
  \includegraphics[width=1.5in]{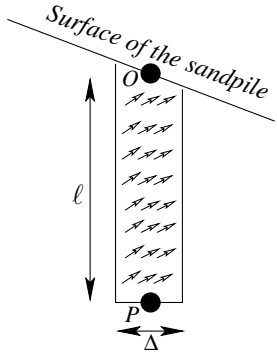}
  \caption{To predict fluidization, we calculate the total seepage
    force on a slice of width $\Delta$ and height $\ell$ and compare
    it with its weight.}
  \label{fig:fluidize}
\end{figure}

Fluidization occurs when at some point $P$ in the sandpile the pore
pressure is larger than the total hydrostatic pressure due to the
weight of the sand and the water above $P$.  To see this we compute
the total seepage force acting on a slice of sand of width $\Delta$
between point $P$ and point $O$ on the surface of the pile directly
above $P$.  The vertical component of this force is (see
\fig\ref{fig:fluidize})
\begin{equation}
  \label{eq:vert_fluid}
  F_\mathrm{v} = \rho g \Delta (\psi_\mathrm{P} -
  \psi_\mathrm{O}) = \rho g \Delta (\pi_\mathrm{P} - \pi_\mathrm{O} -
  \ell),  
\end{equation}
where $\ell$ is the height of the slice.  When this force exceeds the
submerged weight $\Delta \ell (\rho_\mathrm{t} - \rho) g$ of the
slice, the slice is lifted and the bed is fluidized.  Here
$\rho_\mathrm{t}$ is the total density of the saturated sand, which
for our sand is approximately 2 g/cm$^3$.  Thus fluidization occurs
when there exists points $P$ and $O$ on the surface directly above $P$
such that
\begin{equation}
  \label{eq:fluidization}
  \pi_\mathrm{P} - \pi_\mathrm{O} > \ell \rho_\mathrm{t}/\rho.
\end{equation}
For uniform seepage this condition is equivalent to those in
\cite{iverson86:_ground} and \cite{martin71:_seepage}.

To construct the fluidization boundary in the phase diagram
(\fig\ref{fig:phase}), we find the water level $H_\mathrm{fluid}$
above which there exists at least one point in the pile for which
condition~\eqref{eq:fluidization} is satisfied.  Below this
fluidization water level this condition is not satisfied for any point
in the pile.

In addition to fluidization the sandpile can fail by slumping.  This
can happen in one of two ways.  Frictional failure can occur in the
bulk of the pile due to the seepage stresses.  Alternatively, surface
avalanching can occur.  To establish an upper bound on the water level
at which the sandpile slumps via either mechanism we use the criterion
developed by \cite{iverson86:_ground} for determining when a slope is
destabilized by uniform groundwater seepage.  Essentially it requires
calculating the vectorial sum of the seepage and gravity forces acting
on a small element of soil near the surface.  When the angle between
this total force and the downward normal to the surface, which we will
call the effective inclination angle, exceeds the maximum angle of
stability, the surface grains are destabilized.  We measured the
maximum angle of stability to be $\varphi \approx 23^\circ$ for dry
glass beads.  The slumping boundary in the phase diagram
(\fig\ref{fig:phase}) is constructed by computing the effective
inclination angle along the surface of the pile and noting the water
level $H_\mathrm{slump}$, at which the effective inclination angle
reaches the maximum angle of stability $\varphi$ at some point of the
surface.

Figure \ref{fig:phase} shows the critical water level at which
fluidization and slumping should occur according to the criteria
above.  Failure occurs at systematically lower water levels in the
experiment.  There are several effects which can account for this
difference.  First, any irregularities in the construction of the pile
such as voids or surface height fluctuations make the pile more
unstable to fluidization and slumping.  Second, we compute the
instability of an uneroded pile, whereas in most experiments, the pile
failed after erosion had changed the shape of the pile.  The decrease
of pile's height due to erosion increases the head gradient in the
bulk and thus makes the pile more prone to slumping and/or
fluidization.

\begin{figure}[htbp]
  \centering 
  \includegraphics[width=3.3in]{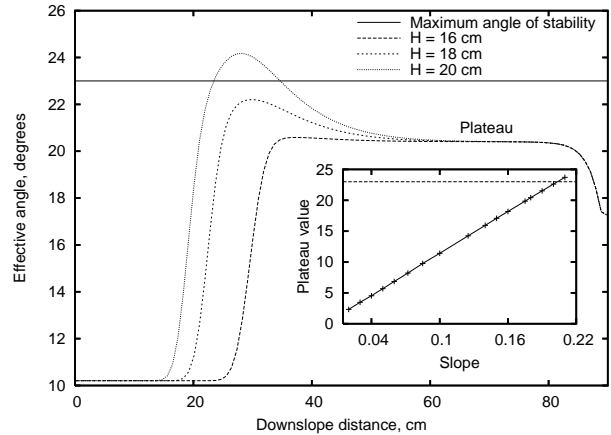}
  \caption{Effective inclination angle at the surface due to seepage
    forces for $s = 0.18$ and three different water levels.  For the
    highest water level, a region on the surface has an effective
    angle above the maximum angle of stability and thus the slope is
    unstable to slumping.  The inset shows the plateau value of the
    effective inclination angle as a function of slope.  When the
    plateau value reaches the maximum angle of stability, even a small
    amount of seepage destabilizes the pile to slumping.}
  \label{fig:slump_jump}
\end{figure}

At $s \approx 0.2$, a jump in the slumping water level is observed in
both the experiment and the model.  This jump is a purely geometric
effect.  Slumping occurs when, somewhere along the slope, the
effective inclination angle, which includes the effect of the seepage
force, exceeds the maximum angle of stability.  As shown in
\fig\ref{fig:slump_jump}, for slopes smaller than $0.2$, the
effective inclination angle is flat and develops a peak under the
water inlet as the water pressure is increased.  When the top of this
peak crosses the value of the maximum angle of stability, the pile
slumps.  When the slope exceeds $0.2$, however, the value of the
plateau in the effective inclination angle is above the maximum angle
of stability.  Therefore, for these slopes, the pile will be unstable
to slumping as soon as the water emerges on the surface.

\section{Conclusions}
\label{sec:conclusions}

This article reports on our progress in understanding seepage erosion
of a simple non-cohesive granular material in a laboratory-scale
experiment introduced in \cite{schorghofer04:_spont}.  Our ultimate
goal is to construct a quantitative predictive theory of the onset and
growth of the channel network observed in this experiment.  This goal
requires a complete sediment transport model as well as the
calculation of the relevant water fluxes.  Here we obtain the latter
and focus on the onset of erosion.

Prediction of the onset of erosion based on the generalized Shields
conjecture explains qualitatively the channelization boundary in the
experimental phase diagram.  By invoking well established simple ideas
we also roughly explain the fluidization and slumping boundaries in
the phase diagram.  Greater discrepancy with the experiment for these
boundaries indicates that better understanding of the
slumping/fluidization mechanisms particular to our experiment is
needed.

The central result of our exploration is the introduction of the
generalized Shields criterion for seepage erosion.  As a consequence
of seepage forces on the surface grains, the threshold for the onset
of erosion driven by seepage is slope dependent.  The threshold
disappears at a critical slope $s_\mathrm{c}$ determined by the
critical Shields number for overland flow and the density contrast
between the granular material and water.  In most cases this critical
slope is significantly smaller than the maximum angle of stability.
We find, therefore, that slopes above this critical slope are unstable
to any amount of seepage.  As a consequence, slopes that sustain
seepage must be inclined at an angle smaller than the critical angle
for seepage erosion.  This behavior contrasts strongly with the
threshold phenomena in erosion by overland flow, and therefore
provides a mechanistic foundation for distinguishing the two types of
erosion.

\section{Acknowledgements}
\label{sec:ack}

This work was supported by a DOE Grants DE-FG02-99ER15004 and
DE-FG02-02ER15367.

We thank N.~Sch\"orghofer, W.~Dietrich, and K.~Whipple for helpful
discussions.

\bibliographystyle{agu}

\end{article}

\end{document}